# Tree-frog-inspired osmocapillary adhesive bonding to diverse substrates

Zefan Shao[1], Rui Ji[1], Qihan Liu[1]*

[1]Department of Mechanical Engineering and Materials Science, University of Pittsburgh; Pittsburgh, 15213, USA.

*Corresponding author. Email: qihan.liu@pitt.edu

**Abstract:** The performance of conventional pressure-sensitive adhesives (PSA) critically relies on the wetting of the adhesive over the substrate. When the wetting is hindered by low surface-energy substrates or the presence of interfacial liquids, the adhesion is greatly compromised. Here, inspired by tree frogs that secrete mucus with surfactants to achieve good wetting over diverse dry and wet substrates, we explore a novel mechanism, osmocapillary adhesion, to use interfacial liquid modified by surfactant to wet diverse substrates and achieve robust reversible adhesion. We show that osmocapillary adhesives greatly outperform conventional PSAs over low-energy, moist, oily, and greasy substrates while achieving similar performance on high-energy surfaces.

Pressure-sensitive adhesive (PSA) is one of the most widely used adhesives for its ease of application: its solid state avoids the messiness of applying liquid glue and its instantaneous adhesion under light pressure avoids the hassle of activation or curing (*1, 2*). Like all adhesives, adhesive-substrate wetting is crucial to ensure intimate contact, which is a prerequisite of good adhesion. PSAs struggle to bond to low-energy substrates that are hard to wet (*3*), or substrates contaminated with interfacial liquids that interrupt adhesive-substrate contact, such as moisture, oil, and grease (*4*). To improve wetting over low-energy surfaces, one can use low-surface-energy polymers to synthesize PSAs, such as silicone and fluorinated polymers (*5, 6*). To improve wetting over moist surfaces, one can add hydrophilic groups into PSA (*7, 8*). However, lowering the surface energy of a PSA increases hydrophobicity, yet making a PSA more hydrophilic increases surface energy. Adhesion over diverse substrates covered by diverse liquids, such as biological tissues covered by aqueous solutions and lipids, can be particularly challenging for conventional PSAs (*9*).

Reversibly adhering to diverse substrates, including low-energy and contaminated surfaces, however, is part of tree frogs' daily life. Unlike conventional PSAs, which are dry, tree frogs secrete mucus to generate capillary adhesion (*10-12*). Since the mucus itself is water-based, surface moisture can be absorbed into the mucus without interfering with the mucus-substrate wetting. Surfactants secreted with mucus enhance wetting over low-energy surfaces. Surfactants can also solubilize oil and grease from the contaminated substrate. Good wetting of the mucus over a substrate leads to concave menisci between the toe pad and the substrate, and the capillary pressure puts the mucus under tension, generating a suction in the mucus (Figure 1A). This type of suction-based adhesion is known as capillary adhesion (*13*). Since tree frogs' adhesion relies on the wetting of an interfacial liquid, it can be easily enhanced by a trace amount of surfactant (*14*). In contrast, conventional PSAs rely on the interaction between the polymer backbone and the substrate. Modifying the polymer backbone to tune wettability often involves complex synthesis and/or complex phase separations (*5-8*). Adding surfactants to conventional PSAs only worsens the adhesion as the interfacial surfactants interrupt the polymer-substrate interaction (*15-17*).

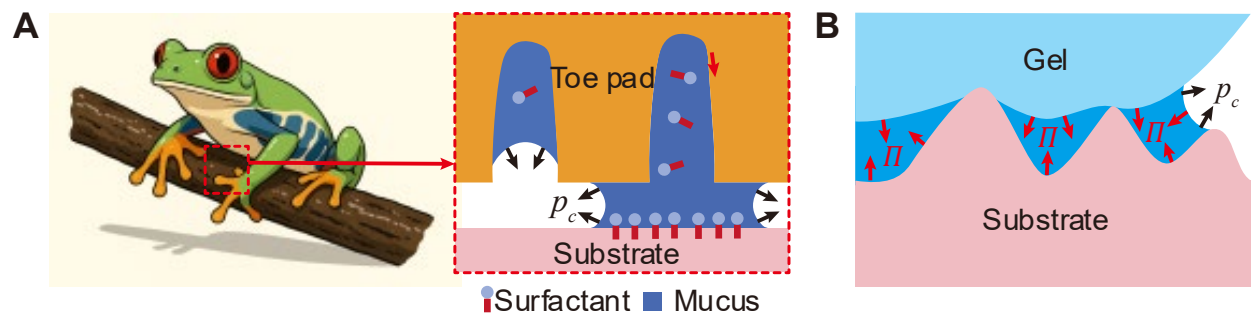

**Fig. 1 Similarity between Tree-frog adhesion and osmocapillary adhesion. (A)** Tree frogs secrete a surfactant-containing mucus to facilitate capillary adhesion. **(B)** On a gel surface, the balance of osmotic pressure $\Pi$ and capillary pressure $p_c$ can lead to osmocapillary phase separation. The osmocapillary phase separation functions similarly to the mucus of the tree frogs in generating adhesion.

Here we use a simple mechanism called osmocapillary phase separation to realize tree-frog-inspired adhesion in any polymer network. Polymer networks tend to absorb favorable solvents and swell into a gel, e.g., hydrogel swells in water and rubber swells in mineral oil. At thermodynamic equilibrium, the gel pulls the solvent with a tensile stress, $\Pi$, known as the osmotic pressure. If the solvent is stress-free, the osmotic pressure drives the continuous absorption of the solvent into the gel until the gel is fully swollen or the solvent is fully consumed. On the gel surface, however, if a concave meniscus is present, capillary pressure $p_c$ can potentially balance the osmotic pressure $\Pi$, leading to a coexisting solvent phase at thermodynamic equilibrium (Figure 1B). This phenomenon is called the osmocapillary phase separation (*18-22*). On a gel-substrate interface, the osmocapillary phase separation works like the secret from tree frog pads, generating a suction of the magnitude of $\Pi$. We call this type of adhesion the osmocapillary adhesion (*23*).

Osmocapillary adhesion is similar to the classical capillary adhesion because both rely on the suction generated by capillary pressure through liquid bridges (*13*). Osmocapillary adhesion differs from classical capillary adhesion in two important ways. First, classical capillary adhesion is caused by the extrinsic presence of interfacial liquids, such as the condensation from humidity or the secretion from animal glands. In contrast, osmocapillary adhesion is an intrinsic material property because the solvent in the osmocapillary phase separation must maintain thermodynamic equilibrium with the solvent in the gel, thus the suction force in the osmocapillary phase separation is defined by the osmotic pressure $\Pi$ in the gel bulk. Second, since capillary pressure is only strong when the meniscus is small, conventional capillary adhesion generally requires highly smooth interfaces (roughness <10nm) or special microstructures (feature size <10μm) for noticeable adhesion (*24, 25*). In contrast, since polymeric gels are highly deformable and osmotic pressure $\Pi$ is easily orders of magnitude higher than a gel's elastic modulus (*20, 23*), osmocapillary adhesion readily conforms to surface roughness and strong adhesion can be achieved without stringent requirements on the surface morphology. These two differences make osmocapillary gel adhesion perform like conventional PSA rather than classical capillary adhesion.

**Surfactant assisting osmocapillary adhesion over diverse substrates**
In an earlier work, we have established the mechanism of osmocapillary adhesion using polyacrylamide (PAAm) hydrogel (*23*). While PAAm lacks the interfacial bonding mechanisms required in the conventional design of hydrogel adhesives or the rheological properties in the conventional design of PSA, we have shown that ~100kPa level tensile self-adhesion strength and ~100J/m$^2$ level self-adhesion energy can be achieved by simply dehydrating the hydrogel, which increases the osmotic pressure $\Pi$. We have further shown that self-adhesion strength does not depend on the elastic modulus or the polymer volume fraction of the gel. This independence indicates that the solid part of the hydrogel has a negligible contribution to adhesion strength. Then the monotonic dependence on the osmotic pressure $\Pi$ indicates that the adhesion is osmocapillary in nature.

Here we extend the study to diverse substrates. We perform flat-probe tack tests to measure the adhesion strength of a 60% v/v PAAm hydrogel over substrates of different surface energies. It is known that a mechanical test only reveals the defect-free strength if the defects present in the system are much smaller than the fractocohesive or fractoadhesive lengths, $\Gamma/W_f$, where $\Gamma$ is the fracture energy or adhesion energy and $W_f$ is the strain energy density to fracture a defect-free sample (*26, 27*). Since the flat-probe tack test minimizes the bulk deformation in the adhesive (Figure S1), thus minimizing $W_f$, the defect-free adhesion strength is more easily measured. We find that the adhesion over a high-energy substrate, glass, is noticeably stronger than over a low-energy substrate, polytetrafluoroethylene (PTFE), indicating the importance of wetting (Figure 2A). This substrate dependence is similar to conventional PSA (*3*). However, while adding surfactant weakens the adhesion of conventional PSA (*15-17*), it strengthens osmocapillary adhesion. The strengthening effect reaches a plateau above the critical micelle concentration (CMC). The same trend is observed on different surfactants and substrates. Here sodium dodecyl sulfate (SDS) is an anionic surfactant and Triton X-100 is a non-ionic surfactant. Since surfactants only reduce solvent surface energy up to CMC (*14*), this plateau suggests that osmocapillary adhesion is dominated by the solvent-substrate wetting. Adding surfactants also reduces the precompression required to reach the maximum adhesion strength compared to conventional PSA (Figure S2). Moreover, there appears to be a master curve between the adhesion strength and the solvent-substrate contact angle when comparing the measurements from different substrates and surfactants (Figure 2B). The existence of a material-independent relation suggests that osmocapillary adhesion is solely governed by the solvent-substrate wetting. The polymer network plays a negligible role. This is distinct from conventional PSAs, where the polymer-substrate wetting is crucial.

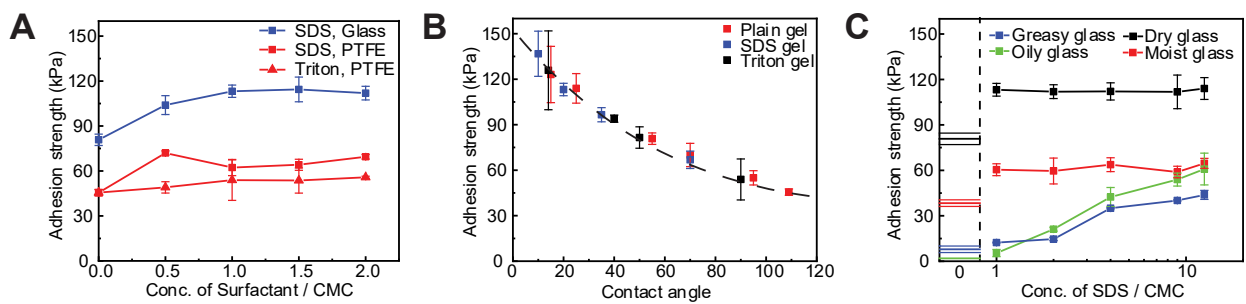

**Fig. 2 Surfactant strengthens osmocapillary adhesion. (A)** Osmocapillary is weaker over a low-energy surface (PTFE), indicating the importance of good wetting. Surfactants can enhance solvent-substrate wetting thus enhancing adhesion strength up to CMC. The CMCs of Triton X-100 and SDS are 0.2mM and 8.0mM (*28, 29*). **(B)** The surfactant effect on adhesion strength over different substrates (identical PAAm hydrogel, glass, nitrile rubber, PTFE, in the order of increasing contact angle) can be fully collapsed by the change in contact angle. Surfactant concentrations are at the corresponding CMCs. The dashed line is a visual trend line. **(C)** Hydrogel readily absorbs surface moisture and maintains adhesion, although the strength is lowered due to the lowered osmotic pressure $\Pi$. Surfactants can solubilize interfacial oil and grease when the concentration is much higher than CMC, recovering the adhesion strength over oily and greasy substrates. Error bars indicate standard deviation, n = 3 for each group.

We next study the osmocapillary adhesion over surfaces contaminated with moisture, oil, and grease. For the moist surface, we use an ultrasonic humidifier to spray a uniform layer of water droplets of μm diameter on a glass substrate (Figure S3). For the oily and greasy surfaces, we coat mineral oil or silicone grease over the glass and remove excess with a razor blade. Different surface preparations are used because water has a finite contact angle on the glass thus forming discontinuous droplets while oil and grease perfectly wet the glass thus forming a continuous film. All three surface contaminations weaken osmocapillary adhesion but show different dependence on the surfactant concentration (Figure 2C). Since PAAm hydrogel readily absorbs interfacial moisture by itself, the adhesion strength over the moist substrate shows the same surfactant dependence as the dry substrate. The adhesion strength is lower because absorbing moist swells the gel near the interface, which locally lowers the osmotic pressure $\Pi$. In contrast, the adhesion strengths over the oily and greasy surfaces nearly vanish without surfactant, indicating that water in the PAAm hydrogel cannot penetrate the oil or grease film to wet the substrate. As the surfactant concentration increases above the CMC, the adhesion strength recovers. This is consistent with the capability of micelles to solubilize oil and grease above CMC (*14*). The removal of the interfacial oil and grease is corroborated by the contact angle measurement before and after the adhesion test (Figure S4). Since the micelles that solubilize interfacial oil and grease are much larger than the water molecules, thus having much slower kinetics, the absorption of interfacial oil and grease is slower than interfacial moisture. It takes about 10s to establish adhesion over the oily and greasy substrate in contrast to less than 1s over dry and moist substrates (Figure S5).

**Independently tuning adhesion strength and adhesion energy**

While the flat probe tack test uniformly separates the bonding surface thus minimizing the deformation in the adhesive, debonding in practice often involves complex deformation in the adhesive. In such cases, bulk dissipation is crucial in resisting the propagation of an interfacial crack (*30, 31*). The effect of bulk dissipation in interfacial crack propagation is characterized by the adhesion energy, which is commonly measured through the peel test (*32*). Lap shear test, while commonly used to measure adhesion strength, rarely characterizes the defect-free strength due to the large $W_f$ induced by the uniform shear. The reported lap shear strength often depends on the sample thickness (*27, 33*) and should be adapted to calculate the adhesion energy (*27*). We have verified that the lap shear strength increases with bulk dissipation and is lower than the probe-tack strength for our samples, indicating that the lap-shear strength is affected by defects. On the other hand, the lap-shear and peel tests result in identical adhesion energies, consistent with the existing study (*27*). In the rest of the paper, we will stay with probe-tack strength and peel adhesion energy.

Since water serves as a plasticizer between polymer chains, PAAm hydrogel generally has low bulk dissipation (Figure S7). Correspondingly, the adhesion energy is low, ~10J/m². We can enhance bulk dissipation by introducing transient entanglements into the polymer network. This strategy does not modify the chemistry of the polymer-solvent system or the polymer-solvent ratio, thus not affecting the osmotic pressure $\Pi$. Then the effect of bulk dissipation can be isolated from interfacial osmocapillary interaction. We increase the number of transient entanglements by reducing the crosslinker-monomer ratio $C$ and the polymer volume fraction during synthesis $\phi_0$, which in turn reduces the number of crosslinks and permanent entanglements in the polymer network (*34, 35*). We found that the adhesion energy increases as the gel becomes more dissipative (Figure 3A). Here resilience characterizes the ratio of the strain energy recovered upon unloading (*31*). The lower the resilience, the higher the dissipation. At the same time, since the gel osmotic pressure and the solvent surface energy remain unchanged, the adhesion strength remains unchanged (Figure 3B). This indicates that the interfacial osmocapillary interaction and the bulk dissipation function independently. We can further increase the number of transient entanglements by mixing uncrosslinked PAAm chains into the gel and reach $\Gamma = 500 \text{J/m}^2$ (Figure S8). However, free chains can be pulled out of the gel, lowering the adhesion strength. Nevertheless, the independence of osmocapillary interaction and bulk dissipation means that one can design osmocapillary adhesives like bond-based adhesives, and the great variety of established bulk dissipation mechanisms can be employed to toughen the adhesion (*31*).

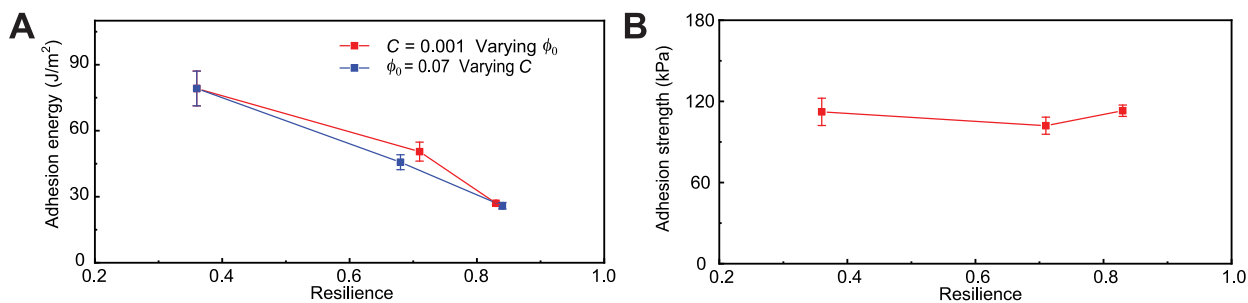

**Fig. 3 Bulk dissipation enhances adhesion energy independent of adhesion strength. (A)** Adhesion energy increases as gel becomes more dissipative. Here a lower resilience corresponds to higher dissipation. **(B)** Adhesion strength is independent of resilience. Error bars indicate standard deviation, n = 3 for each group.

## Generality of osmocapillary adhesion

The mechanism of osmocapillary adhesion can be implemented in any polymer network (Figure 4A). Water in PAAm hydrogel can be replaced with glycerol to make the adhesive nonvolatile. One can also replace PAAm with polyhydroxyethylmethacrylate (PHEMA) and use PEG-400 as the solvent to formulate another nonvolatile adhesive. The same principle applies to non-polar polymer-solvent systems such as polybutadiene-dodecane (BR-dodecane). We fix the polymer volume fraction at 60% v/v for all systems, except for BR-dodecane at 90% v/v. Different polymer-solvent interactions can lead to different osmotic pressure $\Pi$. Consequently, adhesion strengths differ between systems.

Adding surfactants close to CMC noticeably enhances adhesion in all osmocapillary adhesives except that neither SDS nor Tritan X-100 is soluble in dodecane. Here the CMC in glycerol and PEG 400 is estimated by contact angle measurements (*36, 37*) (Figure S9). The osmocapillary adhesives modified by surfactants show less substrate dependence than commercial PSAs. For example, one of the strongest PSA, acrylic-based VHB tape, shows a factor of 4 drop in adhesion strength between the high-energy glass substrate and the low-energy PTFE substrate. While silicone-based PSA are known to perform better over low-surface energy substrates (*1, 5*), a factor of 2 drop between glass and PTFE is still observed on silicone-based APT polyimide tape. In all cases, osmocapillary adhesives outperform commercial PSAs on the low-energy substrates such as PTFE and nitrile rubber.

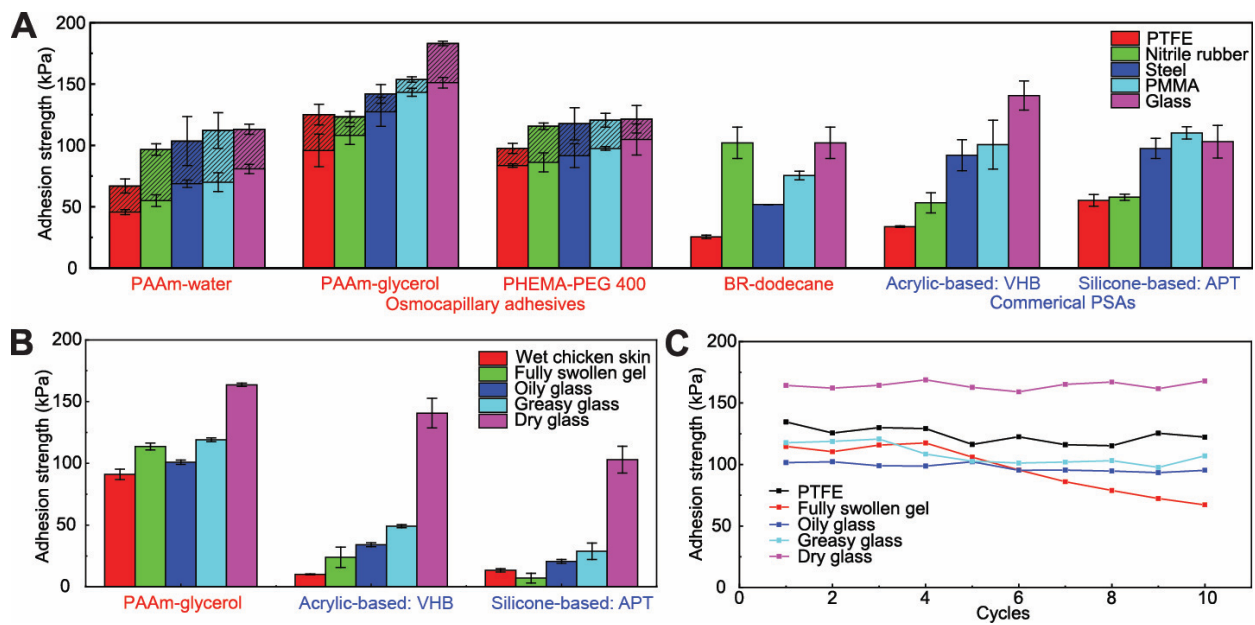

**Fig. 4 Generality and reversibility of osmocapillary adhesion.** **(A)** Various osmocapillary adhesives outperform commercial PSAs on low-energy substrates and show less substrate dependence. The shaded area represents the enhancement by adding 8mM SDS in water or 4mM Triton X-100 in glycerol or PEG 400. **(B)** PAAm–glycerol adhesive with 400mM Triton X-100 maintains robust adhesion on contaminated surfaces (moist, oily, greasy) where commercial PSAs fail. **(C)** PAAm–glycerol adhesive shows reversible adhesion on both hydrophilic and hydrophobic substrates. A noticeable drop in adhesion strength is only observed on the fully swollen gel as water absorption into the adhesive lowers the osmotic pressure $\Pi$ near the surface. Error bars indicate standard deviation, n = 3 for each group.

We then choose one osmocapillary adhesive, PAAm-glycerol, and characterize its adhesion strength over various contaminated substrates (Figure 4B). Besides the oily and greasy glasses studied in Figure 2C, we further include fully swollen hydrogel and wet chicken skin in the comparison. Fully swollen hydrogel has >90% v/v of water and constantly maintains a hydration layer when compressed, leading to an extremely slippery surface (*38*). Animal skin is also challenging to bond to as its surface can be covered with both the oily sebum and aqueous sweat (*39*). We found that commercial PSAs, whether acrylic-based or silicone-based, show more than a factor of 3 drop in adhesion strength over oily and greasy substrates and more than a factor of 6 drop on moist surfaces. In contrast, the osmocapillary adhesive maintains more than half of its dry-glass-strength over all contaminated surfaces.

Osmocapillary adhesion is expected to be highly reversible because interfacial osmocapillary phase separation is a thermodynamic phenomenon that happens reversibly. Indeed, we find that the adhesion strength changes negligible over ten bonding-debonding cycles over dry substrates such as glass and PTFE and contaminated surfaces such as oily and greasy glass (Figure 4C). There is a drop in adhesion strength over moist surfaces, the fully swollen hydrogel, because absorbing the interfacial moisture locally lowers the osmotic pressure $\Pi$ near the interface. Here each bonding-debonding cycle is performed on a different spot on the contaminated substrate. All 10 cycles are finished within 10 minutes.

In summary, osmocapillary adhesion is a general mechanism to generate adhesion in any polymer network. Unlike conventional PSAs, osmocapillary adhesion is dominated by the wettability of the solvent instead of the polymer. By adding surfactants, osmocapillary adhesion can wet low-surface-energy substrates better than conventional PSAs. With sufficiently low interfacial energy, osmocapillary adhesion can realize nearly identical adhesion properties across substrates of diverse surface energies. The combination of hydrophilic solvent and surfactants can further absorb moisture, oil, and grease from the interface, allowing osmocapillary adhesion to work on contaminated substrates. The combination of substrate-insensitive adhesion and contamination removal makes osmocapillary adhesion a universal and robust alternative to the conventional PSAs.

**Acknowledgement:** This work is supported by National Science Foundation under Grant no.2337592.

# Supplemental Materials

## Tree-frog-inspired osmocapillary adhesive bonding to diverse substrates

Zefan Shao[1], Rui Ji[1], Qihan Liu[1]*

[1]Department of Mechanical Engineering and Materials Science, University of Pittsburgh; Pittsburgh, 15213, USA.

*Corresponding author. Email: qihan.liu@pitt.edu

## Materials and Methods

### Synthesis of PAAm-water adhesive

Acrylamide (AAm, Sigma-Aldrich A8887), N,N′-methylenebisacrylamide (MBAA, Sigma-Aldrich 146072), and ammonium persulfate (APS, Sigma-Aldrich A3678), are used as the monomer, crosslinker, and thermal initiator in gel synthesis (all purchased from Sigma-Aldrich and used as received). Deionized (DI) water (McMaster-Carr 3190K731) is used as the solvent. For the baseline formulation ($\phi_0 = 0.51, C = 0.001$), 9.0 g AAm is dissolved in 6.0 g DI water under continuous stirring until fully dissolved. MBAA and APS stock solutions are prepared at 0.15 g/10 mL and 0.1 g/10 mL, respectively. Subsequently, 1302 µL of MBAA stock solution and 231 µL of APS stock solution are added to the monomer solution with gentle stirring. Sodium dodecyl sulfate (SDS, Sigma-Aldrich L4509) or Triton X-100 is optionally added at concentrations ranging from 0 to 100 mM as noted in the manuscript. Custom-built molds are assembled using glass plates (McMaster-Carr 8476K16) and acrylic spacers cut to size with a $CO_2$ laser cutter (XTool P2, 55 W). Each mold consists of a glass substrate, an acrylic spacer, and a glass cover plate. The spacer is positioned on the substrate without adhesives, and the precursor solution—with or without the surfactant, is poured into the mold and sealed with the cover plate. A 500 g calibration weight (Bekith) is placed on top to ensure a tight closure. Polymerization is carried out on a hotplate (Ohaus, guardian 5000) at 60 °C for 3 hours to produce polyacrylamide (PAAm) hydrogel.

To examine bulk dissipation, monomer solutions are prepared at reduced polymer volume fractions ($\phi_0 = 0.29$ and $0.07$) by dissolving 5.6 g AAm in 9.4 g DI water or 1.5 g AAm in 13.5 g DI water, respectively. For $\phi_0 = 0.07$, the crosslinker-monomer ratio $C$ for the sample is varied from 0.001 to 0.0015 and 0.002 using 325 or 434 µL of MBAA stock solution, together with 481 µL of APS stock solution. SDS is added at a concentration of 8mM. Then the precursor solution is polymerized following the same protocol as the baseline formulation. The synthesized hydrogels are dehydrated with a heat gun to reach $\phi = 0.51$, with the target weight calculated from $\phi_0$ following $m_t = 2.5 m_0 \rho_{AAm} / (\phi_0^{-1} - 1 + \rho_{AAm})$, where $m_t$ is the target weight, $m_0$ is the as-synthesized weight, and $\rho_{AAm} = 1.443$ g/cm$^3$ is the density of AAm. This yields PAAm hydrogels of identical polymer fraction but with different bulk dissipation.

### Synthesis of PAAm-water adhesive with uncrosslinked chains

A PAAm solution of uncrosslinked polymer chains is synthesized as follows. A monomer solution is prepared by dissolving 9.0 g AAm in 6.0 g DI water under continuous stirring. Methyl 3-mercaptopropionate (MMP, Sigma-Aldrich 108987) is used as the chain transfer agent to control the average chain length. Irgacure 2959 (Sigma-Aldrich 410896) is used as the photo initiator. Irgacure 2959 stock solution is prepared at 0.1 g/10 mL. Then 337 µL of MMP and 284 µL of Irgacure 2959 stock solution are added to the monomer solution with gentle stirring. SDS is added at a concentration of 8mM. The mixture is thoroughly stirred and cured under UV light for 3 hours to produce uncrosslinked PAAm chains.

The solution of uncrosslinked PAAm chains is subsequently mixed with baseline precursor solution at varying network ratios (0.0–1.0) and polymerized following the same thermal curing protocol, yielding hydrogels with the same total polymer fraction,

Synthesis of PAAm-glycerol adhesive

The same monomer, crosslinker, and thermal initiator used for the PAAm–water adhesive are employed here. A 1:1 (w/w) mixture of DI water and glycerol (Sigma-Aldrich G6279) is used as the solvent to ensure uniform mixing and prevent phase separation during polymerization. To prepare the precursor, 6.0 g of AAm is dissolved in 8.0 g of the solvent mixture under continuous stirring. MBAA and APS stock solutions are prepared at 0.15 g/10 mL and 0.10 g/10 mL, respectively, and 868 µL of MBAA stock solution and 193 µL of APS stock solution are added sequentially with gentle mixing. Triton X-100 is optionally incorporated at concentrations of 0, 4, or 400mM. The precursor solution is cast and polymerized following the same procedure as the PAAm–water adhesive. After curing, the gels are further heated at 60 °C over 12 hours to remove residual water until the weight does not change, yielding the final PAAm-glycerol adhesive.

Synthesis of PHEMA-PEG 400 adhesive

2-hydroxyethyl methacrylate (HEMA, Sigma-Aldrich 128635), poly(ethylene glycol) dimethacrylate (EGDMA, Sigma-Aldrich 335681), and APS are used as the monomer, crosslinker, and thermal initiator, respectively (all purchased from Sigma-Aldrich and used as received). Polyethylene glycol 400 (PEG 400, Sigma-Aldrich 8074850050) is used as the solvent. To prepare the precursor, 9.0 g of HEMA is mixed with 6.0 g of PEG 400 under continuous stirring. APS stock solution is prepared at 0.1 g/10 mL and 13.1 µL of EGDMA and 631 µL of APS stock solution are sequentially added to the monomer–solvent mixture with continuous stirring. Triton X-100 is optionally added at a concentration of 4.0mM. The precursor solution is thoroughly mixed and cured following the same procedure as the pAAm-water adhesive. After curing, the gels are further heated at 60 °C over 12 hours to remove residual water until the weight does not change, yielding the final PHEMA-PEG 400 adhesive.

Synthesis of BR-dodecane adhesive

Polybutadiene (BR, Sigma-Aldrich 181374), 1,6-hexanedithiol (Sigma-Aldrich W349518), and Irgacure 2959 (Sigma-Aldrich 410896) are used as the monomer, crosslinker, and photo initiator, respectively. Tetrahydrofuran (THF, Sigma-Aldrich 360589) served as the solvent. To prepare the precursor solution, 2.0 g of polybutadiene is dissolved in 8.0 mL of THF, followed by the addition of 5.64 µL of 1,6-hexanedithiol and 0.008 g of Irgacure 2959. The mixture is thoroughly stirred and cured under UV light for 3 hours. After curing, the rubber is placed in a fume hood to allow complete solvent evaporation, then swollen in dodecane (Sigma-Aldrich D221104) to achieve a 90% v/v BR content, yielding the final BR–dodecane adhesive.

Moist substrate preparation

A moist substrate is prepared using a humidifier (AquaOasis Cool Mist Humidifier) operating at maximum power, with the nozzle positioned parallel to the glass substrate (McMaster-Carr 8476K16) at a distance of ~5 cm. The substrate is sprayed for 10 s, producing a uniform droplet layer with an average diameter of ~25 μm as confirmed by optical microscopy. Immediately after preparation, the moist substrate is brought into contact with the test sample, and the adhesion test is performed without delay to minimize evaporation.

Oily substrate preparation

An oily substrate is prepared using mineral oil (McMaster-Carr 3025K43). Approximately 5 μL of oil is dispensed onto a glass substrate (McMaster-Carr 8476K16) using a pipette (Eppendorf Research Plus, 0.5-10 μL) and evenly spread over an approximate 3 cm × 5 cm area by gentle finger swiping to ensure complete coverage of the bonding region in the probe–tack test. Excess oil is removed with a clean razor blade, yielding a uniform, thin oil film over the designated area.

Greasy substrate preparation

A greasy substrate is prepared using vacuum grease (Molykote High Vacuum Silicone Grease Tube). Approximately 0.1 g of grease is transferred onto a glass substrate (McMaster-Carr 8476K16) using a toothpick (L'ELEGANCE) and evenly spread over an approximate 3 cm × 5 cm area by gentle finger swiping to ensure full coverage of the bonding region in the probe–tack test. Excess grease is carefully removed with a clean razor blade, yielding a uniform, thin grease layer across the designated area.

Wet chicken skin substrate preparation

A wet chicken skin substrate is prepared using fresh chicken wings (bought from local supermarket Giant Eagle and used within 2 days of storing in the fridge). The wings are first soaked in tap water to remove residual blood, and the skin is carefully separated from the underlying muscle with a clean razor blade. Excess water is blotted with a paper towel, and the skin is stretched flat to a 3 cm × 5 cm area (Fig. 4B). The prepared skin is glued to a glass plate (McMaster-Carr 8476K16) using cyanoacrylate adhesive (Gorilla Super Glue), yielding a uniform, thin, and flat chicken skin substrate for testing.

Fully swollen gel substrate preparation

A fully swollen gel substrate is prepared from a polyacrylamide formulation with an initial polymer volume fraction of $\phi_0 = 0.41$ and a crosslinker-to-monomer ratio of $C = 10^{-4}$. The gel is immersed in DI water for 24 h until its mass does not change. The corresponding polymer volume fraction is $\phi = 0.09$. The swollen gel is then cut into a 3 cm × 5 cm piece and bonded to a glass substrate (McMaster-Carr 8476K16) using cyanoacrylate adhesive (Gorilla Super Glue). The substrate is used within 5 minutes of preparation to avoid dehydration. Immediately before each probe–tack test, 10 µL of DI water is dispensed onto the gel surface to maintain full swelling. Excess water is gently removed by finger swiping, ensuring a uniform, fully swollen surface over the test area.

Flat-probe tack test

Adhesion strength is measured using a flat-probe tack test. Square gel samples (15 mm × 15 mm, ~ 1 mm thickness) or polybutadiene samples (15 mm × 15 mm, ~ 0.1 mm thickness) are bonded to a glass plate (McMaster-Carr 8476K16) using cyanoacrylate adhesive (Gorilla Super Glue). To ensure parallel alignment, the two glass plates to bond the samples and the substrates are first pressed into contact before being fixed to the universal testing machine (Shimadzu EZ-XL-HS). The sample and the substrate are bonded to the fixed glass plate after this aligning procedure. During testing, the gel surface is brought into contact with the substrate at an approach speed of 0.5 mm min$^{-1}$ until reaching the targeted normal preload. The preload is maintained for a defined holding time (1 s, 10 s, or 100 s). The fixtures are then separated at a rate of 1mm min$^{-1}$. The force during the separation is recorded using the 100N load cell. Adhesion strength $s$ is calculated as $s = F_{max}/A$, where $F_{max}$ is the peak force and $A$ is the initial contact area.

Peel test

Rectangular samples (20 mm × 100 mm) of the osmocapillary adhesive are applied to target substrates under a uniform preload applied by rolling a 2 kg metal cylinder (diameter = 7.5 cm, thickness = 4.5 cm). The free surface of the adhesive is then bonded to an inextensible PET backing layer (McMaster-Carr 8567K102) using cyanoacrylate (Gorilla Super Glue). Samples are peeled from the substrates at a fixed 90° angle and a constant rate of 50 mm min$^{-1}$ by a universal testing machine (Shimadzu EZ-XL, 100 N load cell) and the force displacement curve is continuously recorded. Adhesion energy is calculated as $\Gamma = F_{ss}/w$, where $F_{ss}$ is the measured steady-state peeling force (Fig. S6) and $w$ is the sample width.

Lap shear test

Square samples (20 mm × 20 mm) of the osmocapillary adhesive are bonded to glass slides (McMaster-Carr 8476K16) using cyanoacrylate glue (Gorilla Super Glue). The sample is manually pressed onto the substrate with a 5 mm offset (Fig. S6), leaving a 20 mm × 15 mm contact area. An offset larger than the sample thickness is important in getting reliable fracture energy (27), which is calculated as $\Gamma = \frac{h}{2\mu}(\frac{F_{max}}{wL})^2$. Here $h$ is the adhesive thickness, $\mu$ is the shear modulus of adhesive, $F_{max}$ is the maximum force during the lap shear test, and $w$ and $L$ are contact width and length. The sample is tested on a universal testing tester (Shimadzu EZ-XL, 100 N load cell) at 50 mm min$^{-1}$.

Resilience measurements

The resilience of samples is measured using a cyclic pure-shear tensile test. Rectangular samples (45 mm × 5 mm) with thickness ranging from 0.63 to 1.17 mm are tested using a universal tensile tester (Shimadzu EZ-XL) with a 100 N load cell. Here the samples have different thickness due to the adjustment of the polymer volume fraction after the synthesis. Samples are stretched at a constant rate of 10 mm min$^{-1}$ along the loading direction. Each sample is loaded to $\lambda = 2$, and then unloaded back to $\lambda = 1$.

Nominal stress is calculated as $s = F/(wt)$, where $F$ is the loading force, $w = 45\ mm$ is the initial sample width and $t$ is the initial thickness. The restored and dissipated strain energies, $W_R$ and $W_D$ are then integrated from the stress–stretch curves as schematically illustrated in Fig. S7A.

Reversibility test

The reversibility of the adhesives is evaluated using repeated flat-probe tack tests on both clean and contaminated substrates. Clean substrates (PTFE and glass) and contaminated substrates (oily glass, greasy glass, and fully swollen hydrogel) are prepared as previously described. For each sample, ten probe tack tests are performed at ten different locations on the same substrate. Different locations are used to ensure identical surface conditions at each repetition. Adhesion strength for each cycle is determined following the same protocol as the single-cycle probe-tack test, with each measurement representing one cycle.

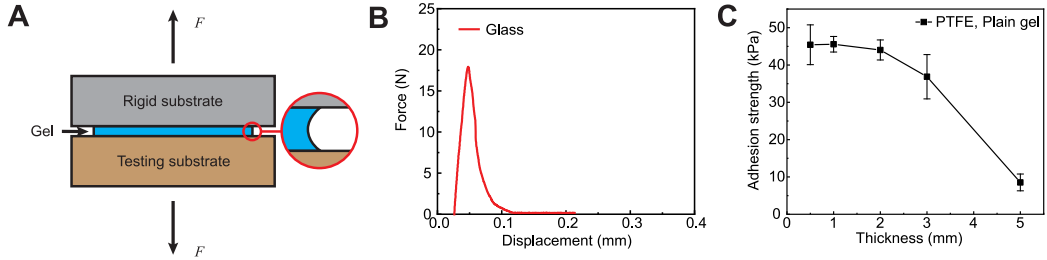

**Fig. S1. Flat-probe tack test.**
**(A)** A thin layer of gel is bonded to a flat rigid substrate and pulled from a flat testing substrate. Since the gel is nearly incompressible, the lateral constraint from the rigid substrate also constrains the deformation in the pulling direction. Consequently, the gel layer experiences no deformation during the test except near the edge, where the lateral constraint is missing. As long as the gel layer is thin and large enough, the effect of the edge deformation is negligible. By minimizing deformation, the flat-probe tack test minimizes $W_f$, thus maximizing $\Gamma/W_f$, which ensures that the strength measured is in the defect-free regime. **(B)** A representative force–displacement curve for a piece of plain gel on a dry glass substrate. The peak force, $F_{max}$, corresponds to the maximum tensile stress sustained by the interface and is used to calculate the adhesion strength. **(C)** The region of inhomogeneous edge deformation scales with sample thickness. For a fixed in-plane sample size, increasing the sample thickness increases the edge deformation thus increasing the average $W_f$. Consequently, the adhesion strength is affected by defects for thick samples, leading to a drop in the strength. We ensure that the adhesion strength reported is in the thickness-independent range, which reflects the uniform separation of the interface. Error bars indicate standard deviation, n = 3 for each group.

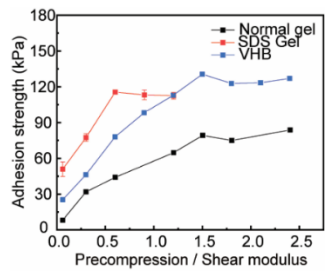

**Fig. S2. Effect of precompression on adhesion strength.**
Adhesion strength increases with the normal precompression in a probe tack test. The adhesion strength reaches a plateau when the precompression is comparable to the elastic modulus of the adhesive. Surfactant-modified gels achieve maximum strength at a lower preload compared to unmodified gels and commercial VHB tape. Here the concentration of the SDS is 0.6 times CMC. Error bars indicate standard deviation, n = 3 for each group.

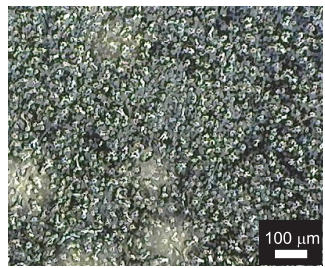

**Fig. S3. Preparation for a moist substrate.**
Water droplets on moist substrates have an average diameter of ~25 μm after 10 s of spraying at a distance of 5 cm from the outlet of an ultrasonic humidifier.

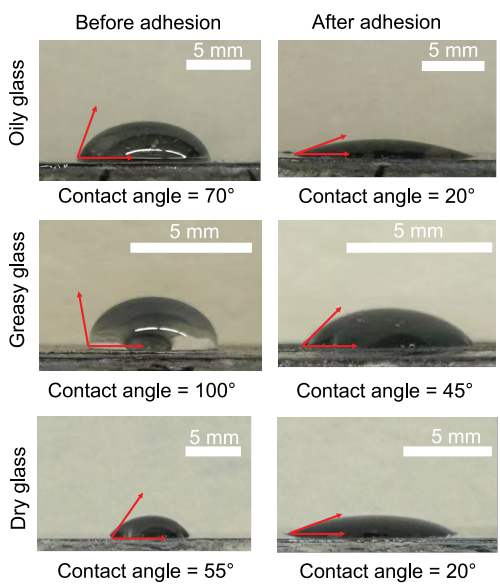

**Fig. S4. Surfactant enhances adhesion on dry, oily and greasy substrates.**
The water contact angles over the oily and greasy substrates noticeably decrease after the adhesion tests with the 100mM-SDS-PAAm hydrogel. The decrease in contact angle indicates the removal of interfacial oil and grease. The contact angle over dry glass is also reduced after adhesion tests, indicating that some residual surfactant is left over the substrate.

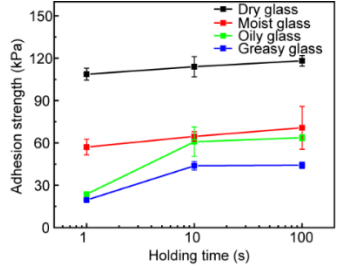

**Fig. S5. Effect of holding time on different substrates.**
The adhesion strength of 100mM-SDS-PAAm-water hydrogel is nearly instantaneous on dry and moist substrates. It takes ~10s to establish the adhesion over oily and greasy substrates. Error bars indicate standard deviation, n = 3 for each group.

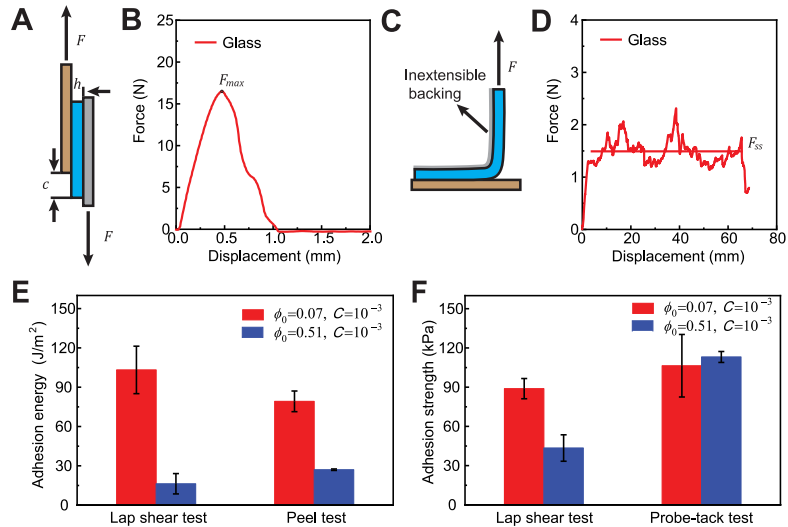

**Fig. S6. Adhesion energy and strength from standard 90° peel and lap shear tests.**
(**A**) Schematic of the lap shear test. (**B**) Representative force–displacement curves of the lap shear test. The adhesion energy is calculated from the peak force $F_{max}$. (**C**) Schematic of the peel test. (**D**) Representative force–displacement curves of the peel test. The adhesion energy is calculated from the steady-state force, $F_{ss}$. (**E**) Both the lap shear test and peel test yield identical adhesion energy within experimental error. Peel tests have smaller error bars because the steady-state force is averaged over a distance. (**F**) The lap shear strength is affected by dissipation and is lower than the probe-tack strength. The probe-tack strength is independent of the bulk dissipation. Error bars indicate standard deviation, n = 3 for each group. Here $\phi_0$ indicates the polymer volume fraction at the synthesis state. $C$ indicates the crosslinker-monomer ratio during synthesis.

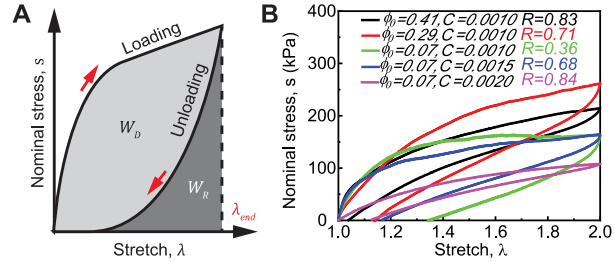

**Fig. S7. Resilience of pAAm-water adhesives.**
**(A)** Schematic of uniaxial loading–unloading cycle used to quantify resilience ($R$), defined as the ratio of recovered energy ($W_R$) to total input energy ($W_R + W_D$), where $W_D$ is the dissipated energy. **(B)** Nominal stress–stretch loops of PAAM–water adhesives with varying network parameters, showing tunable dissipation. $R$ values are extracted from each loop and indicated in the legend.

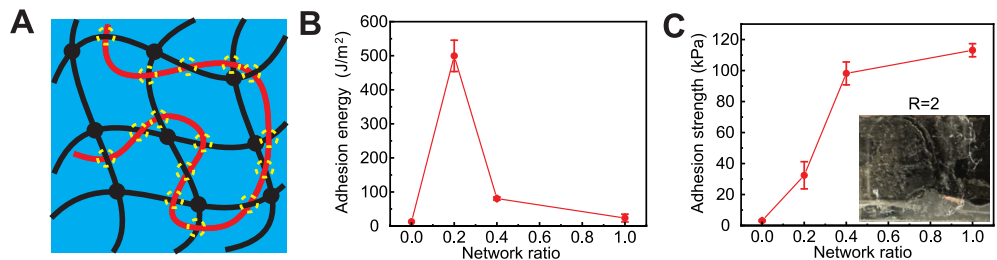

**Fig. S8. Effect of uncrosslinked chains on gel adhesion.**
**(A)** An uncrosslinked chain (red curve) can induce a large number of transient entanglements (yellow dotted circles) in a crosslinked network (black curves are polymer chains, solid black circles are crosslinks). The network ratio is defined as the volume fraction of the crosslinked polymer divided by the total volume of the polymer in the gel. **(B)** We keep the total polymer fraction fixed while mixing the uncrosslinked chains with the crosslinked network. Although the uncrosslinked chain itself (network ratio 0) and the crosslinked network itself (network ratio 1) both have low adhesion energy, mixing them can significantly enhance the adhesion energy. **(C)** Mixing uncrosslinked chains into the network will lead to free chains being pulled out from the gel, leaving residues on the substrate. The adhesion strength decreases as more residue is pulled out. Error bars indicate standard deviation, n = 3 for each group.

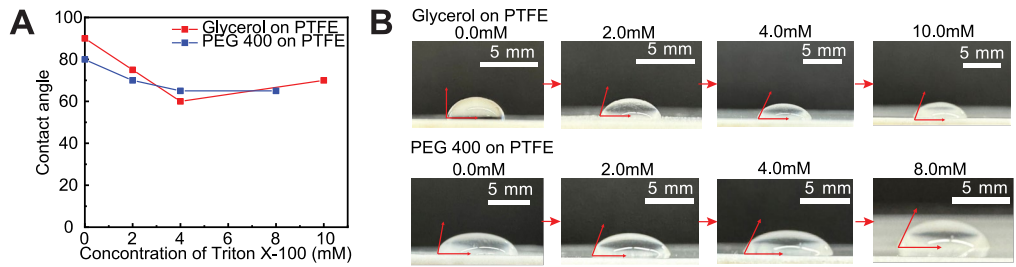

**Fig. S9. Estimation of the critical micelle concentration (CMC) of Triton X-100 in non-aqueous solution.**
**(A)** Contact angles of glycerol and PEG 400 on PTFE decrease with increasing Triton X-100 concentration and reach a plateau around ~4 mM, which corresponds to the CMC. **(B)** Photos of the contact angle measurements for glycerol and PEG 400 on PTFE.